\documentclass[aps,pra,reprint,superscriptaddress,showpacs]{revtex4-1}
\usepackage{amsmath,amssymb,graphicx,hyperref}
\begin{document}
\title{Single photon absorption and dynamic control of a coupled quantum dot-cavity system}

\author{R. Johne}
\affiliation{COBRA Research Institute, Eindhoven University of Technology, PO Box 513, NL-5600MB Eindhoven, The Netherlands}
\author{A. Fiore}
\affiliation{COBRA Research Institute, Eindhoven University of Technology, PO Box 513, NL-5600MB Eindhoven, The Netherlands}

\date{\today}

\begin{abstract}
We theoretically investigate the dynamic interaction of a quantum dot in a nanocavity with time-symmetric single photon pulses. The simulations, based on a wavefunction approach, reveal that almost perfect single photon absorption occurs for quantum dot-cavity systems operating on the edge between strong and weak coupling regime. The computed maximum absorptions probability is close to unity for pulses with a typical length comparable to the half of the Rabi period. Furthermore, the dynamic control of the quantum dot energy via electric fields allows the freezing of the light-matter interaction leaving the quantum dot in its excited state. Shaping of single photon wavepackets by the electric field control is limited by the occurrence of chirping of the single photon pulse. This understanding of the interaction of single photon pulses with the quantum dot-cavity system provides the basis for the development of advanced protocols for quantum information processing in the solid state.    \end{abstract}


\maketitle
\section{Introduction}
Over the past decades it has been proved that quantum communication and quantum computing can provide some advantages compared to their classical counterparts \cite{Ladd2010}. But the question which physical implementation is most suited remains open. The optimal qubit should satisfy two conflicting requirements at the same time: First, it should be perfectly decoupled from the environment because each interaction will disturb the fragile quantum mechanical system and induces decoherence. On the other hand, a controlled interaction between qubits is essential for the building blocks of quantum computers as well as for the manipulation of information in quantum communication network nodes \cite{Briegel1998}. Photons are unchallenged as flying qubits due to weak decoherence and excellent transport properties but the vanishing small photon-photon interaction is strongly disadvantageous.

In a fundamental work about the physical implementation of quantum computing \cite{DiVincenzo2000} the transfer of flying qubits into stationary qubits (e.g. matter qubits), which provide the necessary interaction, has been proposed to overcome the limitations of photonic qubits. This transformation requires a perfect light-matter interface with unity probability emission and absorption of flying qubits. Cavity quantum electrodynamics provides a promising system for such an interface due to the enhanced light matter coupling.

Various systems can provide a stationary two-level system coupled to a resonator mode including atoms \cite{Raimond2001,Leibfried2003} and superconducting qubits \cite{Clarke2008,Schoelkopf2008}. Here, we will focus on a solid-state system,
namely electron-hole pairs (excitons) in quantum dots (QDs). The solid state solution has a particular advantage, because QDs can be integrated in a solid state cavity and they are accessible via waveguides, which enables in principle large-scale integration. Their usefulness as non classical light sources is well established \cite{Shields2007}. The embedding of quantum dots into single and multimode cavities not only enhances the light matter interaction, it can also be used to improve the performance of nonclassical light sources \cite{Dousse2010}  as well as to compensate imperfections \cite{Johne2008,Johne2009}. Furthermore, the solid state implementation opens the opportunity to manipulate the QD exciton energies with electric fields.

Beside the widely studied emission of photonic qubits from quantum systems also the absorption of single photon pulses is of fundamental interest for applications in photonic quantum networks. The ultimate challenge is the perfect quantum state transfer from a source system to a similar target system by means of a single photon. Without further engineering, typical candidates for such network nodes, atoms and quantum dots, emit asymmetric photon pulse envelopes in the time domain with a sharp rise and a slow decay tail. It has been shown that a two-level system in free space behaves like a perfect absorber for pulses which perfectly match the time reversed spatiotemporal initial emitter profile i.e. photon wavepackets with a slow rise and a sharp decay \cite{Stobinska2009}. In the light of these findings, the development of quantum interfaces becomes quite challenging due to the asymmetry of single photon pulse emitted by a source and the reduced absorption in a similar target system. The solution are sources, which emit a time-symmetric single photon pulse. Different approaches in atomic quantum optics have been undertaken in order to produce such symmetric single photon pulse envelopes \cite{Rempe2002,Keller2004} and to allow for efficient quantum state transfer between different quantum network nodes \cite{Cirac1997,Pintosi2008}. While these techniques can be applied to three level ($\Lambda$-type) atomic systems they are difficult to implement in the solid state due to the different level structure.

In this paper we investigate how efficiently quantum state transfer can be realized using symmetric single photon pulses, taking into account the specific tools available in the solid state cavity QED. We study the interaction, in particular the absorption, of symmetric single photon pulses within a coupled QD-cavity system. The underlying theoretical model is described in section II. The results are drawn in section III including also the emission of single photons from the quantum dot-cavity system. For both processes, absorption and emission, the impact of dynamic controlled light-matter coupling via electric fields will be discussed. Finally the paper is closed with a summary and conclusions in section IV. 

\section{Theoretical model}
The model we employ is a two-level system coupled to a cavity mode. The cavity mode itself leaks into a continuum of modes called output field in the following. A possible experimental realization would consist of a quantum dot in a cavity, which is coupled via one mirror to a waveguide (one-sided cavity). Furthermore we consider the possibility to apply an electric field to the QD exciton in order to use the Stark effect \cite{Empedocles1997,Zrenner2002,Patel2010} and to tune the dot in and out of resonance with the cavity mode. The realization of QDs in micropillar and photonic crystal cavities  with electrical contacts for Stark-tuning has been shown experimentally by several groups \cite{Francardi2008,Laucht2009,Chauvin2009,Patel2010,Faraon2010}. The scheme of the system of interest is shown in Fig.1(a) together with the relevant level structure. The non-hermitian Hamiltonian of the system reads (setting $\hbar=1$)
\begin{eqnarray}
\label{Hamiltonian}
 H &&=  {\omega _{c}}{a^ + }a + \left( { {\omega _{_{QD}}}(t) - i\gamma } \right){\sigma ^ + }\sigma  + ig(a{\sigma ^ + } - {a^ + }\sigma ) \\ \nonumber
  &&+ i\sqrt {\frac{\kappa }{{2\pi }}} \int\limits_{\omega _c - {\omega _B}}^{{\omega _c+\omega _B}} {d\omega \left( {{a^ + }b(\omega ) - a{b^ + }(\omega )} \right)}  \\ \nonumber
  &&+ \int\limits_{ \omega _{c}- {\omega _{_B}}}^{{\omega _{c}+\omega _{_B}}} {d\omega \left[ {\omega {b^ + }(\omega )b(\omega )} \right]} ,  
\end{eqnarray}
where $a^+ (a)$, $\sigma^+ (\sigma)$ and $b^+(\omega) (b(\omega))$ are the creation (annihilation) operators for the cavity mode, the two level system and the continuum modes of the waveguide, the latter obeying the standard commutation relation $\left[ {b(\omega ),{b^ + }(\omega ')} \right] = \delta (\omega  - \omega ')$. 
The cavity frequency is given by $\omega_c$ and the exciton energy is given by $\omega_{_{QD}}(t)$. We assume that in the output field only modes within a finite bandwidth $[\omega_{c}-\omega_{_B},\omega_{c}+\omega_{_B}]$ have non-negligible contributions to the dynamics.  The possible time dependence of the exciton energy is already included but for the moment we keep $\omega_{_{QD}}(t)=\omega_{_{QD}}$ constant. Finally, $\gamma$ is the decay rate of the quantum dot exciton, $g$ is the coupling between cavity and exciton and $\kappa$ denotes the cavity decay into the waveguide, where we have used the flat-band condition \cite{walls} representing a uniform coupling to all output field modes. The QD decay $\gamma$ is neglected in the following, but it is still displayed in the equations for completeness. Indeed we suppose that the system is operated at temperatures such that the decoherence time is much longer (typically 100s of ps) than the time of the calculated dynamics. Including the decay term the system dynamics can be obtained by using a wavefunction approach for dissipative quantum systems \cite{Dalibard1992}, which is equivalent to Master equation based techniques.

\begin{figure}
\includegraphics[width=1\linewidth]{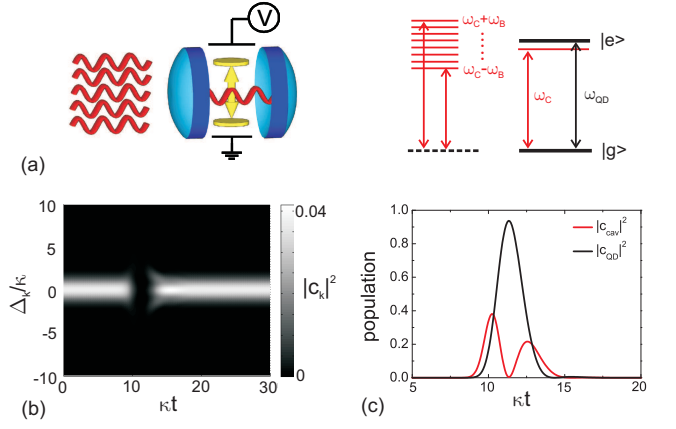}
\label{fig1}
\caption{Schematic of the system: (a) A quantum dot couples via the cavity to a quasi continuum of modes (left) and the relevant energy levels of the system (right). (b) Population of the output field modes $\left|c_k \right|^2$ versus time and detuning. (c) Inner cavity dynamics: The black line corresponds to the QD exciton population and the red line correspond to the cavity mode population. The simulation parameters are $g=\kappa,\gamma=0,w=1/g$.}
\end{figure}

The first line of the Hamiltonian Eq.(\ref{Hamiltonian}) describes the free evolution of the cavity, of the quantum dot and the coupling between these two. The second line describes the coupling between the cavity mode and the continuum in the waveguide. The last term governs the free evolution of the modes of the continuum. We assume that the main decay channel of the cavity is the leakage into the output field and we neglect all other possible decay channels of the cavity. This assumption is justified in high-Q photonic crystal cavities coupled to waveguides \cite{Kim2004}. 

In order to solve the above equation numerically we employ a discretization of the output field ensuring that the frequency range $2\omega_{_B}$ is much larger and the spacing between the output field modes is much smaller than the cavity linewidth.
The Hamiltonian in a rotating frame reads

\begin{eqnarray}
\label{Hamiltonian2}
 H &&= {\Delta _{c}}{a^ + }a + \left( {{\Delta _{_{QD}}}(t) - i\gamma } \right){\sigma ^ + }\sigma  + ig(a{\sigma ^ + } - {a^ + }\sigma ) \\ \nonumber
  &&+ i\kappa '\sum\limits_{k = 1}^N {\left( {{a^ + }{b_k} - ab_k^ + } \right)}  + \sum\limits_{k = 1}^N {{\Delta_k }b_k^ + {b_k}}, 
\end{eqnarray}
where the output field coupling $\kappa'=\sqrt {\frac{{\kappa \Delta \omega }}{{2\pi }}}$. The number of discretized output modes N is set to 1024 in the simulations. The frequency spacing between the quasi-continuum modes is denoted by $\Delta\omega$. The values $\Delta_{_{QD}},\Delta_{cav}$ and $\Delta_k$ are the energy detunings of the two-level transition, the cavity and the output mode $k$ from the rotating frame.

We use a wavefunction approach \cite{Cirac1997,Duan2003,Duan2004} to simulate the dynamics. We expand the wavefunction of the system in all possible states limiting ourselves to the case of a single excitation:
\begin{eqnarray}
\label{wavefunction}
\left| \Psi  \right\rangle && = \left[ {{c_{cav}}\left| g \right\rangle \left| 1 \right\rangle  + {c_{_{QD}}}\left| e \right\rangle \left| 0 \right\rangle } \right]  \left| {vac} \right\rangle  \\ \nonumber
&&+ \left| g \right\rangle \left| 0 \right\rangle   \sum\limits_{k = 1}^N {{c_k}} b_k^ + \left| {vac} \right\rangle .
\end{eqnarray}
The state $\left| {vac} \right\rangle$ denotes the vacuum state of all modes in the quasi-continuum of the waveguide.
The modulus square of the amplitude $c_{cav}$ describes the probability to find one photon in the cavity, $c_{_{QD}}$ is the amplitude for the quantum dot in the excited state and $c_k$ are the amplitudes for the modes of the quasi continuum, satisfying, in case of $\gamma=0$ 
\begin{equation}
{\left| {{c_{cav}}} \right|^2} + {\left| {{c_{_{QD}}}} \right|^2} + \sum\limits_{k = 1}^N {{{\left| {{c_k}} \right|}^2}}  = 1.
\end{equation}
Plugging this wavefunction expansion into the time dependent Schr\"odinger equation $i{\partial _t}\left| \Psi  \right\rangle  = H\left| \Psi  \right\rangle $ yields a system of coupled differential equations, which govern the time evolution of the state amplitudes:
\begin{eqnarray}
\label{System}
 {{\dot c}_{cav}} &&=  - i{\Delta _{c}}{c_{cav}} - {g}{c_{_{QD}}} + \kappa' \sum\limits_{j = 1}^N {{c_k}}  \\ 
 {{\dot c}_{_{QD}}} &&= {g}{c_{cav}} + \left( { - i\Delta _{_{QD}}(t) - \gamma } \right){c_{_{QD}}} \\ 
 {{\dot c}_k} &&=  - i{\Delta _k}{c_k} - \kappa' {c_{cav}} .
\end{eqnarray}

The coupled quantum dot-cavity system interacts with single photon pulses. After a time $T$ when the time evolution is completed the pulse shape $f_{out}(t)$ and the output mode amplitudes are connected via the inverse Fourier transform:
\begin{equation}
{f_{out}}(t) = \frac{1}{{\sqrt {2\pi } }}\sum\limits_{k = 1}^N { {c_k}(T){e^{ - i\omega_k (t - T)}}}
\end{equation}

In a similar way it is possible to drive the system with a single photon pulse by defining an arbitrary pulse shape in the time domain. The Fourier transform of this pulse gives the initial values for the amplitudes $c_k$. The amplitudes have to be properly normalized $\sum\limits_{k = 1}^N {{{\left| {{c_k(t=0)}} \right|}^2}}  = 1$ to ensure that only a single photon is incident. We restrict ourself in the following to Gaussian single photon pulses with the pulse central frequency resonant with the cavity mode and temporal properties given by the width $w$ and $t_0$
\begin{equation}
{f_{in}}(t) \propto {e^{\frac{{^{ - {{(t - {t_0})}^2}}}}{{{2w^2}}}}}.
\end{equation}
In the adiabatic limit, when the pulse amplitude is varying slowly enough in comparison to the decay rate $\kappa$, the input pulse remains almost unperturbed by the interaction with the two-level system \cite{Duan2004}. In contrast, the present study deals with pulse lengths comparable to the decay time of the cavity in order to achieve maximum absorption. Furthermore we assume that a symmetric photon pulse is provided by an independent source where the temporal width $w$ can be chosen freely. The issue of generating such a symmetric pulses from the solid state system is discussed later.

The present theoretical description can be easily modified to account for different geometries e.g coupling to a second waveguide. Also the extension to two photons may be possible but it requires a more complicated analytical description and more demanding numerics.

\section{Results and Discussion}
\subsection{Single photon absorption}

\begin{figure}
\includegraphics[width=0.8\linewidth]{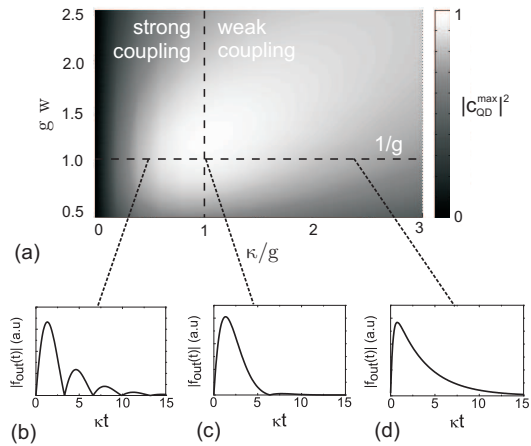}
\label{fig2}
\caption{(a) Maximum quantum dot population $|c_{_{QD}}^{max}|^2$ during the time evolution versus cavity decay $\kappa$ and pulse length $w$ of the incident photon pulse for $g=$const. The vertical dashed line shows the transition from the strong coupling regime (left) to the weak coupling regime (right) while the horizontal line corresponds to the half of the Rabi period ($= 1/g$). The initial emitter profiles $|f_{out}(t)|$ are shown in the strong coupling regime (b) on the edge between strong and weak coupling regime (c) and in the weak coupling regime (d).}
\end{figure}

In the following we want to shed light on the interaction of single photons with the coupled QD-cavity system and, we specifically calculate the efficiency of the single photon absorption by the QD.
The simulations are illustrated in Fig.1(b) and (c). The system is operated on the edge between the strong and weak coupling regime i.e. $g=\kappa$ and we neglect the decay of the exciton by setting $\gamma=0$. We start with a Gaussian single photon pulse in the output field as initial condition. The Fourier transform of $f_{in}(t)$ also yields a Gaussian. The temporal evolution shows that after some time the output field population ${\left| {{c_k}} \right|^2}$ goes almost to zero and the pulse enters into the cavity. 
The inner cavity dynamics are shown in Fig.1(c). The calculation displays that the photon pulse is transferred into the quantum dot state and via a temporal excitation of the cavity mode emitted back into the output field. 

The pulse injection into the cavity depends on the pulse central frequency and the specific parameters of the system. In case of weak coupling, a large detuning of the pulse central frequency compared to the cavity mode will result in reflection of the pulse. On the other hand, in the strong coupling regime, the eigenstates of the system are given by the typical Rabi doublet with peaks shifted by $\pm g$ from the initial cavity frequency (assuming dot and cavity in resonance). Consequently, the pulse only enters into the cavity if the pulse central frequency is close to the frequency of one of these eigenstates. 

The efficiency of the single photon absorption by the quantum dot exciton is given by the maximum of the exciton population for a one-sided cavity system $|c_{_{QD}}^{max}|^2=max(|c_{_{QD}}(t)|^2)$. If and only if the population goes to 1 the system undergoes unity probability absorption of one photon pulse at time T
\begin{equation}
\left| g \right\rangle \left| 0 \right\rangle \sum\limits_{k = 1}^N {c_k b_k^ + \left| {vac} \right\rangle }  \to \left| e \right\rangle \left| 0 \right\rangle \left| {vac} \right\rangle.
\end{equation}

\begin{figure}
\includegraphics[width=1\linewidth]{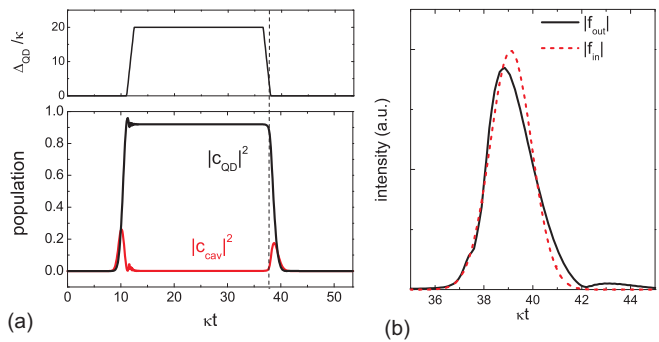}
\label{fig3}
\caption{Dynamic control of the light matter coupling: (a) QD-detuning versus time (upper panel) and inner cavity dynamics (lower panel), where the black line corresponds to the exciton state population $|c_{_{QD}}|^2$ and the red line corresponds to the cavity mode population $|c_{cav}|^2$. (b) Emitted pulse $f_{out}(t)$ with the time shifted incident pulse $f_{in}(t)$. The parameters are taken as $g=\kappa,\gamma=0$ and $w=1/g$. }
\end{figure}

Fig.2(a) displays the maximum exciton population during the interaction process illustrated in Fig.1 (b) and (c) versus the cavity output field coupling $\kappa$ and the incident pulse length $w$. The calculations reveal that there is a maximum absorption probability of about $0.97$ around a pulse length $w=1/g$ and a cavity-output field coupling of $\kappa=g$. The vertical dashed line shows the transition from the strong (left) to the weak coupling regime (right) and the horizontal dashed line shows the half of the Rabi period ($1/g$). The maximum absorption occurs around the crossing of these two lines. For $g$ values typical of InAs quantum dots in photonic crystal cavities ($g \approx 85$ $\mu eV$ \cite{Yoshie2004}), this corresponds to a pulse length of $w\approx8ps$.

The obtained behavior of the quantum dot absorption can be understood in terms of the time reversal symmetry. It has been shown, that a quantum dot in free space behaves like a perfect absorber if the incident pulse perfectly matches  the time reversed spatiotemporal emitter profile \cite{Stobinska2009}. 
In the present case, due to the waveguide coupling, spatial matching is automatically ensured, which is different from the free-space scenario. 
A closer look to the emission properties of the system in the different regimes allows us to shed some light on the observed behavior. We calculate the temporal shape of the emitted pulse envelope (emitter profile) with the above system of equations starting with one exciton in the quantum dot as initial state $|c_{_{QD}}|^2=1$. In the strong coupling regime the emitter profile reveals Rabi oscillations in time domain as shown in Fig.2(b).  The comparison between the incident Gaussian single photon pulse and the emitter profile shows a small overlap. Similar, in the weak coupling regime (Fig.2(d)) the comparison shows a strong mismatch due to the long decay tail of the emitted pulse. In contrast, on the edge between strong and weak coupling regime the emitter emission (Fig.2(c)) profile almost perfectly matches  the incident symmetric Gaussian pulse due to the presence of the cavity coupling and the resulting increase of spontaneous emission (Purcell effect). The absorption is robust against small changes of the system parameters indicated by the rather broad maximum in Fig.2(a). 

An analytical estimate of the maximum absorption value can be obtained by calculating the overlap integral of the incident pulse $f_{in}$ and the initial emitter profile $f_0$:
\begin{equation}
A = \frac{{{{|{\int {{f_{in}^*}(t){f_{0}}(t)dt} }|}^2}}}{{\int {|f_{in}(t)|^2dt\int {|f_{0}(t)|^2dt} } }}.
\end{equation} 
Using the calculated emitter profile and the Gaussian input pulse for the parameters at the maximum absorption on Fig.2(a) and maximizing the overlap integral by shifting the time axis we obtain $A=0.97$, which exactly corresponds to the numerical result.

The results show clearly that the absorption and the emission process are inextricably linked due to the time reversal symmetry of the system.

\subsection{Photon storage using dynamic Stark tuning}

As one can see in Fig.1(c) the quantum dot remains only for a very short time in its excited state due to the coherent light matter interaction. This makes the manipulation of the absorbed qubit very challenging. Furthermore, the preparation of the quantum dot in its excited state in order to obtain a photon-photon nonlinearity mediated by the quantum dot is very challenging because of the short timescale. In order to keep the quantum dot in its excited state dynamic Stark tuning can provide a solution, which is unique to solid-state systems. Assuming that the shift of the quantum dot energies happens instantaneously with the applied field, the quantum dot can be tuned out of resonance once it has reached the maximum state amplitude. Furthermore, after some time shorter than the dipole decay time $\gamma$ (typically in the order of $100s$ of ps) it can be tuned back into resonance and the stored photon can be released. 

The simulations of the dynamic control of the light matter coupling are shown in Fig.3(a) where we display the time dependent detuning of the quantum dot(upper panel) and the inner cavity dynamics (lower panel) when a Gaussian single photon pulse ($w=1/g$) is send to the cavity ($\kappa=g$). Right after the absorption, the quantum dot is shifted out of the cavity resonance within timescales of the order of $t<1/\kappa$. Still some small dynamic features occur in the cavity and quantum dot amplitude but most of the energy can be stored in the QD and the dot cavity interaction can be switched off before it is transferred back to the field. After some time the quantum dot is brought back in resonance with the cavity mode and the stored photon is released.

The released pulse and its comparison to the time-shifted incident pulse are shown in Fig.3(b). Their pulse envelopes are in good agreement because the emitter profile in the strong Purcell regime is close to symmetric. The complete process, absorption and release, can be done with a fidelity of $\approx0.94$ ($\approx0.97^2$), mainly limited by the reduced absorption and the pulse envelope mismatch. For a perfect source the process should have unity fidelity. 

The main constraint for the realization of the dynamic control is the fast change of the electric field. The present systems undergo dynamics in the timescale of tens of $ps$, and its control requires electrical bandwidth above 50GHz, which is feasible but challenging with available electronics. In order to shift the light matter dynamics into a more easily manageable temporal scale one has to use cavities with higher quality factors and quantum dots with smaller cavity coupling $g$. This would allow to slow down the system dynamics to the lower GHz range, where electrical control is easier.

\subsection{Shaping of the single photon pulse}

\begin{figure}
\includegraphics[width=1.0\linewidth]{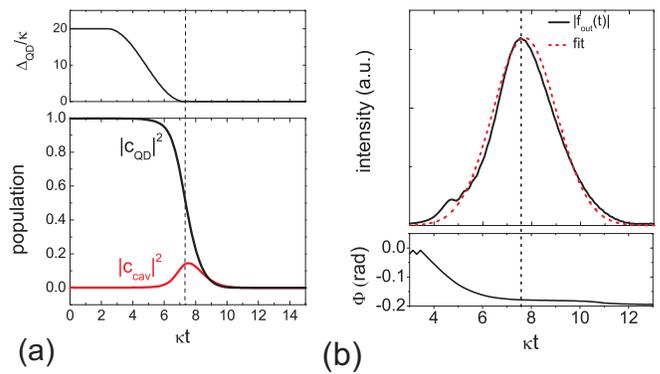}
\label{fig4}
\caption{Shaping of the photon wavepacket: (a) detuning of the quantum dot (upper panel) and inner cavity dynamics (lower panel). The transition from nonzero to zero detuning is indicated by the dashed line. (b) Emitted single photon pulse (black) with a Gaussian fit (red) and the phase $\Phi$ of the pulse shape $f_{out}(t)$ versus time. The parameters are chosen the same as in Fig.3.}
\end{figure}

After the previous discussion of the absorption process we now investigate the possibility of dynamically engineering the quantum dot-cavity interaction in order to store and to produce a time-symmetric photon pulse with unity fidelity. 
In the previous discussion, the Stark tuning is used to abruptly switch on and off the light matter interaction in the system. By changing the detuning on a timescale $t\approx1/g$ it can be also applied to shape the wavepacket of the single photon pulse leaking out of the cavity. This has been proposed to produce symmetric photon pulse envelopes \cite{Fernee2007}. In the present case, the technique is used to shape the first part of photon pulse during the emission process, which is governed mainly by the photon-exciton coupling constant g. By dynamically changing the quantum dot energy the effective coupling to the cavity mode can be manipulated and thus the pulse shape can be controlled.  The starting point for the simulation is one exciton in the far detuned QD. After some time the electric field is applied and shifts the quantum dot into resonance with the cavity. By carefully adjusting the slope of the electric field tuning the emitted pulse shape can be engineered to be remarkably close to a Gaussian envelope in time domain as shown in Fig.4(a) and (b). In both, the transition from finite to zero detuning is indicated by the dashed line. 

The resulting emitter looks at a first glance as a perfect single photon source which emits symmetric single photon pulses. Such a source would also have the the advantage that the absorption of the pulse by a similar system would be possible with unity probability enabling highly efficient quantum state transfer between two nodes \cite{Cirac1997}. Unfortunately, as one can see in figure Fig.4(a), the quantum dot already starts to emit into the cavity mode even when the detuning is slightly different from zero. The resulting output mode population becomes asymmetric in the frequency space and the corresponding output pulse $f_{out}(t)$ has a time-varying phase as shown in the lower panel of Fig. 4(b). 

Using the pulse envelope shown in Fig.4(b) as incident pulse in the simulations yield a maximum absorption $|c_{_{QD}}^{max}|^2<0.8$ due to the frequency chirp and the maximum absorption of $0.97$ obtained in the previous section can not be improved.

One may try to engineer the pulse in such a way, that the frequency chirp becomes time-symmetric. The rising part of the photon pulse is determined by the coupling constant between the exciton and the quantum dot $g$ while the decay is mainly  governed by the complex interplay of QD-cavity coupling g and the leakage out of the cavity $\kappa$. By changing dynamically the detuning and thus the effective coupling of exciton and cavity during the first part of the emission process ($\kappa t<7.5$ in Fig 4.(a)) the quantum dot-cavity interaction can be slowed down. This results in a symmetric pulse envelope in time domain but asymmetric in frequency space as shown in Fig.4. 

On the other hand, the symmetric frequency chirp requirement implies $\Delta_{_{QD}}(t)$ to be time symmetric around $t=7.5/\kappa$.  This condition immediately breaks, at least for the used parameters in Fig. 4, the symmetry of $|f_{out}|$ and the decay part is significantly slowed down in comparison to the rising part of the pulse due to the complex interplay of detuning induced reduced effective exciton-cavity coupling and cavity decay at these times.
The photon pulses in Fig.3 (b) and Fig. 4(b) show the two extreme cases where either the time envelope (Fig.4) or frequency distribution (Fig.3) is symmetric. Extensive numerical studies suggest that the emission of a perfect symmetric pulse is not possible by dynamic Stark tuning of the quantum dot. Nevertheless, further studies are required to prove this conclusion.

Dynamic Q-tuning \cite{Tanabe2006,Tanabe2009} of the cavity together with the application of electric fields might be more complex tools to achieve a perfect pulse but it is not in the scope of the present paper.

\section{Summary and conclusions}

To summarize, we have shown that a coupled quantum dot cavity system acts as an almost perfect absorber for Gaussian single photon pulses when the system is operated on the edge between weak and strong coupling regime. Under this condition the incident pulse closely matches the time inverted initial emitter profile. Furthermore we have shown that the dynamic control of the quantum dot energies can be used as an attractive tool for efficient quantum state preparation. Shaping of single photon wavepackets by the electric field control is limited by the occurrence of chirping of the single photon pulse preventing applications in quantum state transfer protocols. The fast electric field control represents an additional degree of freedom in order to deterministically engineer the light matter interaction in the solid state system.

\subsection{Acknowledgement}
This research is supported by the Dutch Technology
Foundation STW, applied science division of NWO, the
Technology Program of the Ministry of Economic Affairs
under project No. 10380 and from the FOM project No. 09PR2675.

%

\end{document}